\documentclass[twocolumn,prb,showpacs,amsmath,amssymb,english]{revtex4-1}
\usepackage{graphicx}
\usepackage{dcolumn}
\usepackage{bm}
\begin{document}
\title{Resonant and crossover phenomena in a multiband superconductor tuning the
chemical potential near a band edge}
\author{Davide Innocenti$^1$}
\author{Nicola Poccia$^1$}
\author{Alessandro Ricci$^1$}
\author{Antonio Valletta$^2$}
\author{Sergio Caprara$^1$}
\author{Andrea Perali$^3$}
\author{Antonio Bianconi$^1$}

\affiliation{$^1$Physics Department, Sapienza University of Rome,
Piazzale Aldo Moro 2, 00185 Rome, Italy}
\date{\today}
\affiliation{$^2$Institute for Microelectronics and Microsystems, IMM CNR, Via del
Fosso del Cavaliere 100, 00133 Roma, Italy}
\affiliation{$^3$School of Pharmacy, Physics Unit, University of Camerino, 62032
Camerino, Italy}

\date{\today}
\begin{abstract}
Resonances in the superconducting properties, in a regime of crossover from BCS to
mixed Bose-Fermi superconductivity, are investigated in a two-band superconductor where
the chemical potential is tuned near the band edge of the second mini-band generated by
quantum confinement effects. The shape resonances at $T=0$ in the superconducting
gaps (belonging to the class of Feshbach-like resonances) is manifested by interference effects
in the superconducting gap at the first large Fermi surface when the chemical potential is in the
proximity of the band edge of the second mini-band. The case of a superlattice of
quantum wells is considered and the amplification of the superconducting gaps at the 3D-2D Fermi
surface topological transition is clearly shown. The results are found to be in good agreement
with available experimental data on a superlattice of honeycomb boron layers intercalated by Al and
Mg spacer layers.
\end{abstract}


\maketitle

\section{Introduction}

The main physical properties of the Bardeen-Cooper-Schrieffer - Bose-Einstein crossover (BCS-BEC crossover)
at zero temperature for a system of 3D fermions interacting via a contact pairing interaction
can be described with continuity by the BCS (mean-field) equations. Indeed,
as shown by Leggett,\cite{Leggett} the ground-state BCS wave function corresponds
to an ensemble of overlapping Cooper pairs at weak coupling (BCS regime) and
evolves to molecular (non-overlapping) pairs with bosonic character as the
pairing strength increases (BEC regime). The crucial point is that the BCS
equation for the superconducting gap has to be coupled to the equation that fixes the fermion
density: with increasing coupling (or decreasing density), the chemical potential
$\mu$ results strongly renormalized with respect to the Fermi energy $E_F$
of the non interacting system, and approaches $-\frac{1}{2}E_B$, where $E_B$ is
the molecular binding energy of the corresponding two-body problem in the vacuum.
In recent years, the experimental realization of the BCS-BEC crossover in
trapped ultracold fermion gases allowed for a quantitative test of the validity
of increasingly elaborated theoretical approaches: at resonance (the so
called ``unitarity limit''), where the lack of a small parameter does not permit
the systematic control of the approximations, the BCS theory at zero temperature gives a gap
and a chemical potential which are overestimated by about $30\%$ as compared with
the results of QMC calculations and the experimental findings, while already the inclusion of
pair fluctuations at the (non-self-consistent) t-matrix level improves the
comparison for the gap and $\mu$, with deviations not larger than
few percents, as shown by quantitative comparison between theoretical predictions
and experimental results for the BCS-BEC crossover.\cite{Perali04} The mean field approach at
zero temperature, within the local density approximation,\cite{PeraliPRA2003} or in the form of
Bogoliubov-de Gennes equations,\cite{GiorginiRMP2008} has been widely
used in the context of ultracold atoms to study nonuniform configurations
both in homogeneous space and in the confining potential of the trap,
as, e.g., quantized vortices. On the basis of the wide range of quite successful
applications discussed above, in this paper we shall apply the BCS approach (with the
inclusion of the density equation) to the case of multiband superconductivity
with the chemical potential tuned near a band edge. This approach, which has the advantage
of being numerically feasible, is also apt to give a good overall insight into the
resonant and crossover phenomena which are the object of the present work.

Multigap superconductivity in a system with many mini-bands of different
symmetry has been studied in electronic systems with relevant quantum size effects: single
slab,\cite{30,601} single tube,\cite{31} single wire,\cite{34,32} or single quantum dots.\cite{kresin}
The electronic structure generated by the quantum confinement effects gives rise to electron
wave-functions at the Fermi level near a band edge that are strongly affected by the detail of the
quantum confinement. Recently, experimental evidence has been accumulated for superconductivity modulated
by quantum confinement effects in a single quantum well and in a single quantum
dot.\cite{guo,bose1,Zhang,brun,bose2}

Also superlattices of quantum wires or quantum wells show multiband superconductivity near a band
edge.\cite{patent,france,SSC,15d,15f} The three dimensional (3D) structure of the superlattice has the
advantage to suppress quantum fluctuations, that reduce the critical temperature in low dimensions,
while keeping the key features of quantum confinement effects for superconductivity. While there
are works on the theory of shape resonances in two dimensional (2D) superlattices of quantum
wires,\cite{35,18,15g} called ``superstripes'' in cuprate superconductors,\cite{35a,SS,nature}
the theoretical investigation of shape resonances in superlattices of quantum wells is missing. These
are now of high interest since a 3D superlattice of quantum wells provide the simplest case of a 3D
system showing multiband superconductivity near a band edge. Moreover, it has been recognized that
diborides \cite{15a}, intercalated graphite \cite{graf1,graf2,graf3} and pnictides \cite{15h,15h1}
are practical realizations of a superlattice of superconducting layers at atomic limit, where
interband pairing is an essential ingredient for high-temperature
superconductivity.\cite{patent,france,SSC,15d,15f}

In this paper, we study the BCS-BEC crossover which occurs in a two-band system, when the chemical
potential $\mu$ is tuned in a narrow energy range around the edge $E_{edge}$ of the second mini-band.
This description applies to the situation when the first two mini-bands produced by quantum size
effects in a superlattice of quantum wells are well separated, i.e., when the electron hopping between
wells is small enough, to that the corresponding transversal band dispersion $\xi$ is smaller than
the energy separation between the mini-bands.

In Fig. 1 we show that a new 3D Fermi surface (FS) opens when $\mu$ is increased above $E_{edge}$,
and the electron gas in the metallic phase undergoes an electronic topological transition (ETT), also
called a Lifshitz quantum phase transition, of the type ``appearing or disappearing of a new Fermi
surface spot''.\cite{54,15d,15f} When $\mu$ reaches a higher energy threshold, $E_{3D-2D}$, the
electronic structure of the superlattice undergoes a second ETT, the 3D-2D ETT where one FS changes
topology from 3D to 2D (i.e., from ``spherical'' to ``cylindrical'') or vice versa, called ``the opening
or closing of a neck in a tubular FS''.\cite{54,15a,15d,15f} This ETT is typical of i) stacks of metallic
layers, ii)  multilayers, or iii) superlattices of quantum wells, and therefore it is a common
feature of all existing high-temperature superconductors and novel materials synthesized by material
design in the search for room temperature superconductivity. In practical cases, the tuning of $\mu$ at
an ETT can be controlled by means of: i) pressure; ii) the misfit strain\cite{56,57,58} between
the superconducting layers and the spacers; iii) the ordering of dopants in the spacers\cite{nature}
with related striped superstructures; iv) the effective charge density, via gate voltage
methods;\cite{86,87,88} v) the charge transfer between the superconducting layers and
the spacers; vi) the thickness of the spacers.

\begin{figure}[tpb]
\centering
\includegraphics [angle=0,scale=0.35]{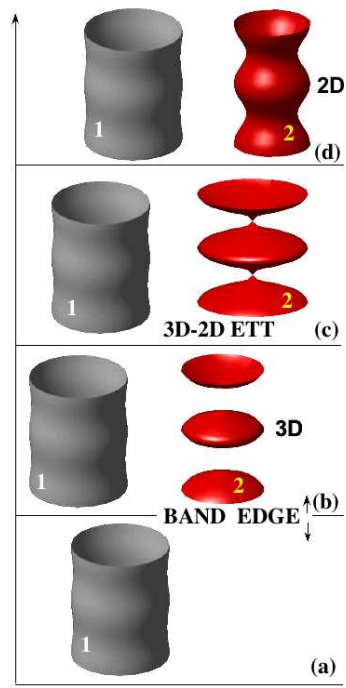}
\caption{Pictorial view of the evolution of the Fermi surfaces (FS) for the superconductor made of non-hybridized two-bands by moving the chemical potential from panel (a) to panel (d) so that it crosses two Lifshitz electronic topological transitions (ETT). The first ETT occurs moving $E_F$ across the band edge $E_{edge}$ of the second band so that the superconductivity goes from the single FS in panel (a) with a single condensate to the two FS in panel (b) with two condensates: the first (1) has a 2D topology and the second (2) has a 3D topology. Here, $\xi$ is the transversal band dispersion.
Changing $E_F$ the system crosses the critical energy $E_{3D-2D}$ where the second FS undergoes a 3D-2D ETT shown in panel (b) changing its topology: the second closed 3D FS [panel (b)] becomes the tubular 2D FS in panel (d). The first large 2D FS (1) remains nearly constant when the chemical potential is moved.}\label{fig1}
\end{figure}
As stated above, we focus on the superconducting properties of a system in which the chemical potential
is varied in a narrow energy range around the band edge, within a simple model for a two-band
superconductor with a first and a second band of different (even and odd) parity and different
topology, which can be varied by tuning the chemical potential. This system provides a particular case
of modulated topological multiband superconductor with multiple different winding
numbers.\cite{topological} We consider the set of coupled BCS equations to study the evolution of
the properties of this two-band system of fermions with contact pairing interactions and a pairing
energy cutoff $\omega_0$ on the order of magnitude of the distance between $\mu$ and the bottom of
the second band. Moreover, resonant and crossover phenomena which are the object of this work,
are amplified when the transversal dispersion $\xi$ is of the same order of magnitude of
$\omega_0$. Therefore, in the following discussion, we concentrate our analysis on systems
that are engineered by material science techniques in such a way that $\xi=\omega_0$.

The condensate in the first large 2D FS, in the standard BCS approximation ($E_F/\omega_0\gg 1$,
i.e., $E_F$ far from band edges), coexists with a second small FS, where the standard
approximation breaks down [$(E_F-E_{edge})/\omega_0\ll 1$, i.e., $E_F$ close to the
band edge where a new FS disappears or appears, or near the 3D-2D ETT, at the closing or
opening of a neck in a tubular FS\cite{54}].

When the chemical potential is located near the band edge, it may be driven below the
bottom of the conduction band by superconducting correlations, in a way similar to the
case of semimetals.\cite{eagles} The phenomenon of quantum shape
resonances\cite{patent,france,SSC,15d,15f} occurs when the chemical potential falls within an
energy range $\omega_0$ near an ETT point. The shape resonances are due to the configuration
interaction between paired states in the large FS, with a large group velocity, and quasi
stationary states near the ETT in the second small FS. The shape resonances in superconducting
gaps near an ETT are analogous to the scattering resonances due to configuration interaction
between a closed and an open channel described in nuclear physics by Majorana and Feshbach, and
in atomic physics by Fano, as reported in a recent work.\cite{vittorini}

In our system, we capture the crucial ingredients of the interplay of the two bands and of the two
set of wave functions (which will be shown to be responsible for the shape resonances in the gap
and in the critical temperature), without facing the complications
beyond BCS theories (as the Eliashberg theory, or corrections beyond Migdal
theorem, etc.) which are considered in more conventional (phonon- or
magnon-mediated) strong-coupling superconductors, or the t-matrix
corrections. These latter would be the
dominant corrections in the regime of band-edge crossing, in particular
for the evaluation of the critical temperature, but should not qualitatively
change our results. Indeed, the Migdal-Eliashberg approach to evaluate the
superconducting properties taking into account the frequency dependence of
the two-particle interaction (retardation effects) can be applied only when
the Fermi energy is much larger than the typical energy of the interaction,
i.e., when the Migdal theorem holds. In multiband superconductors, when
approaching a band bottom, the two energy scales become comparable and
the Migdal theorem cannot be applied (a small parameter controlling the
diagrammatic expansion does not exist anymore), requiring the inclusion of
much more complicated diagrams, as vertex corrections, in determining the
self-consistent equations for the superconducting properties,\cite{PeraliPRB1998}
which is beyond the scope of our work.

We point out that, in this paper, we shall not deal with the problem of superconducting
fluctuations in multiband systems, and refer to the different regimes depending on the position
of the Fermi energy only. Moreover, we recall that a pure BEC limit with the
condensation of point-like bosons, cannot be achieved when a pairing interaction
with a finite energy cutoff is considered in the effective Hamiltonian.
When the Fermi energy crosses both the band edge, the condensate in the newly disappearing or
appearing FS undergoes a crossover from a mixed Bose-Fermi regime to the BCS regime. Indeed,
in the range $-1<(E_F-E_{edge})/\omega_0<0$, the electron states
associated with the newly appearing FS are unoccupied in the normal state, and, in the
presence of a strong enough pairing interaction, a condensate of boson-like tightly bound
pairs is formed below $T_c$. However, we call this regime the mixed Bose-Fermi regime, because
the tightly bound pairs pairs are never genuinely boson-like, due to the presence of the large band.
A preliminary study of the superconducting fluctuations\cite{cu} reveal that these have a mixed
character, propagating (like in the BEC regime) and damped (like in the BCS regime).

The first unconventional BCS regime occurs in the range $0<(E_F-E_{edge})/\omega_0<1$, where all the
few electrons in the newly appearing mini-band, condense below $T_c$. Indeed, in
the case under discussion, $\xi=\omega_0$, the standard approximation
$(E_F-E_{edge})/\omega_0\gg 1$ breaks down. In this range, the new appearing FS has a 3D topology
while the large FS has a 2D topology.

A second unconventional regime occurs around the 3D-2D ETT where the superconductivity phase in
the range $0<(E_F-E_{edge})/\omega_0<2$ arises because of the configuration interaction between
pairing channels in different bands with different condensate symmetry 2D and 3D (i.e., with
different winding numbers) above and below the 3D-2D ETT.

Finally the system is in the standard multiband BCS regime for
$(E_F-E_{edge})/\omega_0>2$.\cite{22,23,21}

There is an analogy between the crossover case studied here and the
BEC-BCS crossover in ultracold Fermi gases. While in the ultracold gases the
tuning of the energy of the bound state of the diatomic molecule above or below the
continuum is performed by using an external magnetic field, here, the ETT in the
second narrow mini-band can be tuned for example by gate voltage techniques,
pressure or misfit-strain, inducing shape resonances in the superconducting
properties.

Here, starting from the standard two-band BCS superconductivity in the clean
limit\cite{22,23,21} far from ETT's, we focus on the case when the chemical potential falls
within a narrow energy range near a band edge where the Feshbach-like \emph{shape resonance}
in superconducting gaps takes place.\cite{15d} The shape quantum resonance in the exchange-like
interband interactionis is generically driven by the repulsive Coulomb interaction and in a
two-band model leads to condensate wavefunctions with opposite signs (a classical quantum
phenomenon, recently becoming popular in pnictides with the name of
$s\pm$ pairing).\cite{16,16a,15d,15f} We show that the control of the ratio between the intensity
of exchange-like interband pairing and intraband Cooper pairing, by material design techniques,
is crucial. We discuss the case of different attractive intraband coupling strengths
in the first and second mini-band and we determine the evolution of the
superconducting gaps by changing the interband exchange-like pairing in the crossover
regime from the BCS regime, well above the band edge to the mixed Bose-Fermi regime,
below the band edge.

We find that the direct evidence for the quantum interference effects between
pairing channels is provided by minima in the gap parameter for electrons in the large FS
and we show the shift to these minima by changing the interband pairing strength.
We report evidence for two different regimes where in the first the interband pairing
dominates while in the second the intraband dominates. These regimes are separated by a critical
value of the interband pairing where the gap ratio becomes independent on the variation
of the chemical potential. Finally, we are able to compare the calculations with the
available experimental data for the evolution of the gaps near the band edge of a 2D FS
in doped diborides, where intraband dominates, but interband pairing is essential for
understanding the evolution of the gaps with the (doping dependent) critical temperature.
The comparison with available experimental data for doped diborides
in a wide range of doping provides a good test of the present theory.

\section{The superconducting gaps}

In a multiband system, near the band edge in the $\ell$-th band, where
$\nabla_{\mathbf k}E_\ell=0$, the energy of an electron can be approximated by
a free-electron dispersion law $E_{\ell,\mathbf k}=E_{\ell}+(\mathbf k^2/2m_\ell)$, where
$m_\ell$ is the electron effective mass at the band edge. However, within our model for
a superlattice of quantum wells disposed along the $z$ axis, while
this approximation is valid to describe the electron dispersion in the $x,y$ plane of the
superconducting layers, it is certainly not valid in the $z$ direction, when the
transversal band dispersion $\xi$ is on the order of energy cut off of the pairing interaction
$\omega_0$. In this case, an anisotropic band with weak dispersion in the $z$ direction and
larger dispersion in the $x,y$ plane above the band edge of the second band should be
adopted,
\begin{equation}
E_{2,\mathbf k}^{3D}=E_{2,L}+E({k_z})+\frac{k_x^2+k_y^2}{2m_\parallel}
\label{1}
\end{equation}
where $E(k_z)$ is the actual energy dispersion in the periodic potential of the
superlattice and $m_\parallel$ is the effective mass in the $x,y$ plane.
This situation is obtained within a model where a free electron gas is confined in
a potential which is periodic in the $z$ direction,
\begin{equation}
\mathcal{W}(z)=\sum_{n=-\infty}^{n=+\infty}\mathcal{W}_b(z-nd),
\label{2}
\end{equation}
where $\mathcal{W}_b(z)=-V_b$ for $|z|\le L/2$ and $\mathcal{W}_b(z)=0$ for
$L/2<|z|<d/2$, $L$ is the width of the confining well and $d$ is the periodicity
of the superlattice in the $z$ direction. This periodic potential mimics the
phenomenology, e.g., of the pnictides, diborides, and stacks of graphene layers
made of a superlattice of stacked planes. The confining potential generates
a band structure organized in mini-bands. Each mini-band has a 3D character,
with closed isoenergetic surfaces, near the lower band edge $E_{edge}=E_{\ell,L}$,
and a 2D character, with isoenergetic surfaces open in the $z$ direction, above
some energy threshold $E_{3D-2D}=E_{\ell,T}$. The present model is apt to describe
quantum interference phenomena between different scattering channels in a large
and a small FS which are the object of this work, when $\mu$ is tuned near the
bottom of a $\ell$-th mini-band, within a window of width $4\omega_0$. In the
energy window of width $2\omega_0$ the first band, with a large 2D tubular FS, has
a constant density of states (DOS) $N_1$. The second FS disappears (or appears) as
$E_F$ crosses the level
$E_{edge}=E_{\ell,L}$. The second FS changes from tubular 2D to closed 3D topology
as $E_F$ crosses the level $E_{3D-2D}=E_{\ell,T}$, as it is shown in Fig. 1. Near
the edge, the DOS of the second FS, $N_2$, has the typical 3D behavior shown
in Fig. 2a in the corresponding range of charge density in the metallic layers
between $\rho_{edge}$ and $\rho_{3D-2D}=\rho_{edge}+\rho_c$.

The crossover from 2D to 3D can be described in our model by changing from an
infinite potential barrier between the planes of the superlattice in the $z$ direction,
provided by spacer layers, to a finite barrier ($V_b$). This yields a finite hopping term
that broadens the sharp discontinuity of the DOS of a pure 2D band. This broadening
increases the width of the shape resonance. Moreover, it is possible to design
artificial superlattice heterostructures at optimum shape resonance condition,
i.e., where the value of the potential barrier $V_b$ and its width are such that
the mini-band dispersion in $z$ direction $\xi$ is of the order of energy cutoff
$\omega_0$ of the interaction, as we shall assume in the following.

It must be pointed out that, differently from the case of a single 2D slab,\cite{30} in
our superlattice,\cite{35} at finite $V_b$ a genuinely 3D condensate is formed, reducing
the effect of fluctuations of the superconducting order parameter, which
suppress the mean-field value of $T_c$. Moreover, coexistence of small and large Cooper
pairs due to different couplings in the two bands is expected to induce an effective
screening of the strong superconducting fluctuations associated with
the small pairs.\cite{Perali2gap}

\begin{figure}[tbp]
\centering
\includegraphics [angle=0,scale=0.50]{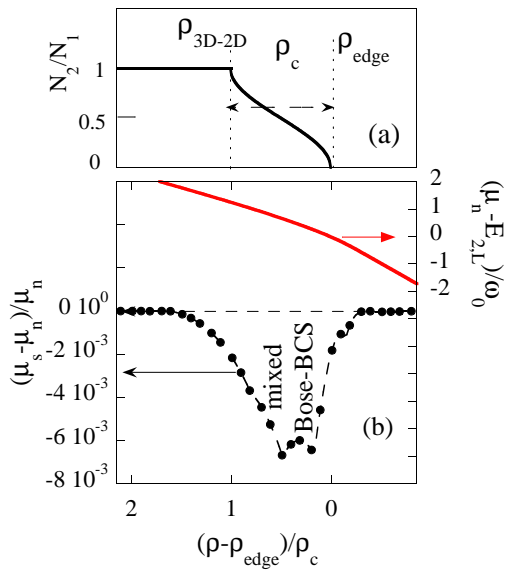}
\caption{(Color online). The crossover regime reached by decreasing the charge
density in the superlattice from the two-band BCS superconductor to the
mixed Bose-Fermi regime at the lover band edge $E_{2,L}$, as a function
of the electron charge density $(\rho-\rho_{edge})/\rho_c$, where $\rho_{edge}$
($\rho_{edge}+\rho_c$) is the charge density where $E_F$ is tuned at $E_{edge}$
($E_{3D-2D}$). Panel (a). The ratio of the DOS of the second mini-band $N_2$ and
the DOS of the first mini-band $N_1$. Upper part of panel (b): The relative variation
of the Lifshitz parameter $(\mu_{n}-E_{2,L})/\xi$ as a function of electron charge;
lower part panel (b): the variation of the chemical potential, $(\mu_{s}-\mu_{n})/\mu_{n}$,
as a function of charge density $\rho$, where $\mu_{n}$ and $\mu_{s}$ are the chemical
potentials in the normal and superconducting phase, $E_{2,L}$ is the lower band edge energy
of the second mini-band.}\label{fig2}
\end{figure}

\begin{figure}[tbp]
\centering
\includegraphics [angle=0,scale=0.50]{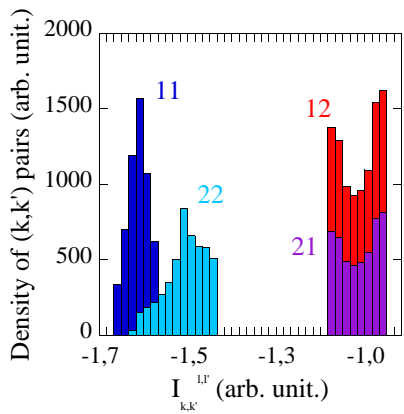}
\caption{(Color online). Density histograms of the pairing interaction matrix
elements $I_{\mathbf{k},\mathbf{k}'}^{\ell,\ell'}$ between the first and the second
mini-bands, for the case with mini-band dispersion $\xi=\omega_0$.
Histograms 11 (blue), 22 (light blue), 21 (violet) and 12 (red) shows respectively
the pairing interaction matrix elements $I_{\mathbf{k},\mathbf{k}'}^{1,1}$,
$I_{\mathbf{k},\mathbf{k}'}^{2,2}$,
$I_{\mathbf{k},\mathbf{k}'}^{2,1}$ and $I_{\mathbf{k},\mathbf{k}'}^{1,2}$.}\label{fig3}
\end{figure}

The pairing interaction is assumed to be originated from an electron-electron contact interaction.
Once the matrix elements between exact eigenstates of the superlattice,
$\widetilde{\mathcal{V}}_{\mathbf{k},\mathbf{k}'}^{\ell,\ell'}$, are calculated they turn out to
depend on the wave vector $\mathbf{k}_z$ in the superlattice direction. This induces a structure
in the $\mathbf{k}$-dependent interband coupling interaction for the electrons that determines
the quantum interference between electron pairs wave functions in different mini-bands of
the superlattice.\cite{15d,15f} The pairing interaction is then cut off at a characteristic energy
$\omega_0$,
\begin{equation}
\mathcal{V}_{\mathbf{k},\mathbf{k}'}^{\ell,\ell'}=
\widetilde{
\mathcal{V}}_{\mathbf{k},\mathbf{k}'}^{\ell,\ell'}
\theta(\omega_0-|\xi_{\ell,\mathbf{k}}|)\theta(\omega_0-|\xi_{\ell',\mathbf{k}'}|)
\label{3}
\end{equation}
where $\mathbf{k}=\mathbf{k}_z$ ($\mathbf{k}'=\mathbf{k}'_z$) is the superlattice
wavevector, in the $z$ direction, perpendicular to the planes, of the initial (final)
state in the pairing process, and
\begin{equation}
\widetilde{
\mathcal{V}}_{\mathbf{k},\mathbf{k}'}^{\ell,\ell'}=
\frac{c_{\ell,\ell'}}{N_0(E_F)V_{3D}}I_{\mathbf{k},\mathbf{k}'}^{\ell,\ell'},
\label{4}
\end{equation}
where $N_0(E_F)$ is the DOS at $E_F$ for a free electron 3D system, $V_{3D}$ is the volume of
the system,
\begin{equation}
I_{\mathbf{k},\mathbf{k}'}^{\ell,\ell'}=-d\int_{d}
\psi_{\ell,-\mathbf{k}}(z)\psi_{\ell',-\mathbf{k}'}(z)
\psi_{\ell,\mathbf{k}}(z)\psi_{\ell',\mathbf{k}'}(z)dz
\label{5}
\end{equation}
and $\psi_{\ell,\mathbf{k}}(z)$ are the eigenfunctions in the superlattice of
quantum wells, normalized so that
$\int_{d}dz|\psi_{\ell,\mathbf{k}}(z)|^2=1$.
The use of single cutoff in two-band superconductors has been justified in detail
by Entel.\cite{Entel} Here, we assume that pairing in our confined
geometry is provided by a contact interaction with a given characteristic energy
range. The dimensionless coupling constants
$c_{\ell,\ell'}=(2\delta_{\ell,\ell'}-1)c^0_{\ell,\ell'}$
are positive for $\ell=\ell'$ (intraband Cooper pairing) and negative
for $\ell\neq\ell'$ (repulsive exchange-like interband pairing, with
$c_{\ell,\ell'}=c_{\ell',\ell}$) and measure the relative intensity
of intraband and interband pairing.

Here a comment on the effective interaction adopted in the present work
is in order. Our pairing interaction is semi-phenomenological in the sense
that we do not derive the microscopic expression of the contact pairing
interaction (its strength and energy range being used as fitting parameters),
but the matrix elements are calculated between eigenstates of the actual
confining potential, and in this sense they are microscopic. The choice
of $\omega_0$ on the order of the mini-band width $\xi$ is justified because this
corresponds to the optimum amplification case, as above discussed.

In order to determine the gaps and $\mu$
self-consistently at zero temperature we use iterative
solving methods for the BCS-like equations
\begin{equation}
\Delta_{\ell,\mathbf{k}}=-\frac{1}{2M}\sum_{\ell',\mathbf{k}'}
\frac{\mathcal{V}_{\mathbf{k},\mathbf{k}'}^{\ell,\ell'}\Delta_{\ell',\mathbf{k}'}}{
\sqrt{(E_{\ell',\mathbf{k}'}-\mu)^2+\Delta_{\ell',\mathbf{k}'}^2}},
\label{6}
\end{equation}
\begin{equation}
\rho=\frac{1}{M d^2}\sum_{\ell,\mathbf{k}}\left[1-
\frac{E_{\ell,\mathbf{k}}-\mu}{\sqrt{(E_{\ell,\mathbf{k}}-\mu)^2+
\Delta_{\ell,\mathbf{k}}^2}}\right],
\label{7}
\end{equation}
where $M$ is the total number of wave-vectors $\mathbf{k}'$ and $\rho$ is the electron
density, starting with initial constant gaps and an initial $\mu$ equal to the value
of $E_F$ in the normal state. We considering convergence to have occurred when the
relative variation of the gap and charge density $\rho$ less than $10^{-6}$.
We point out that the gap functions $\Delta_{\ell,\mathbf{k}}$
depend on the superlattice wave-vector $\mathbf{k}$ as well as
on the mini-band index $\ell$. The BCS-like equations self-consistently couple the gap
function at a given
point of $\mathbf{k}$-space with the values of the gap function in the entire
$\mathbf{k}$-space. In the following we discuss the properties of the average values of the gap
in $\mathbf{k}$-space, unless otherwise specified. The above equations can be easily generalized
to finite temperatures $T$, but in the remaining part of the present section we focus on the
case $T=0$.

\begin{figure}[tbp]
\centering
\includegraphics [angle=0,scale=0.45]{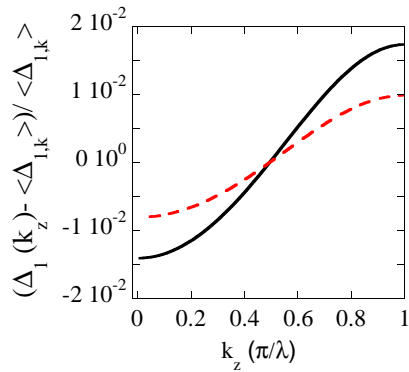}
\caption{(Color online). Calculation of the gap anisotropy, i.e., the dependence of the gap
in the first mini-band, with $c_{22}/c_{11}=2.17$, $c_{11}=0.22$ and $c_{12}/c_{11}=0.43$
as function of the wavevector $k_z$ in the transversal direction. We show in the figure two
characteristics cases with different values of the averaged gap obtained solving the BCS-like
equations at finite temperatures $T=5.7$K (solid line) and $T=27.2$K
(dashed line) in the first mini-band.}\label{fig4}
\end{figure}

\begin{figure}[tbp]
\centering
\includegraphics [angle=0,scale=0.45]{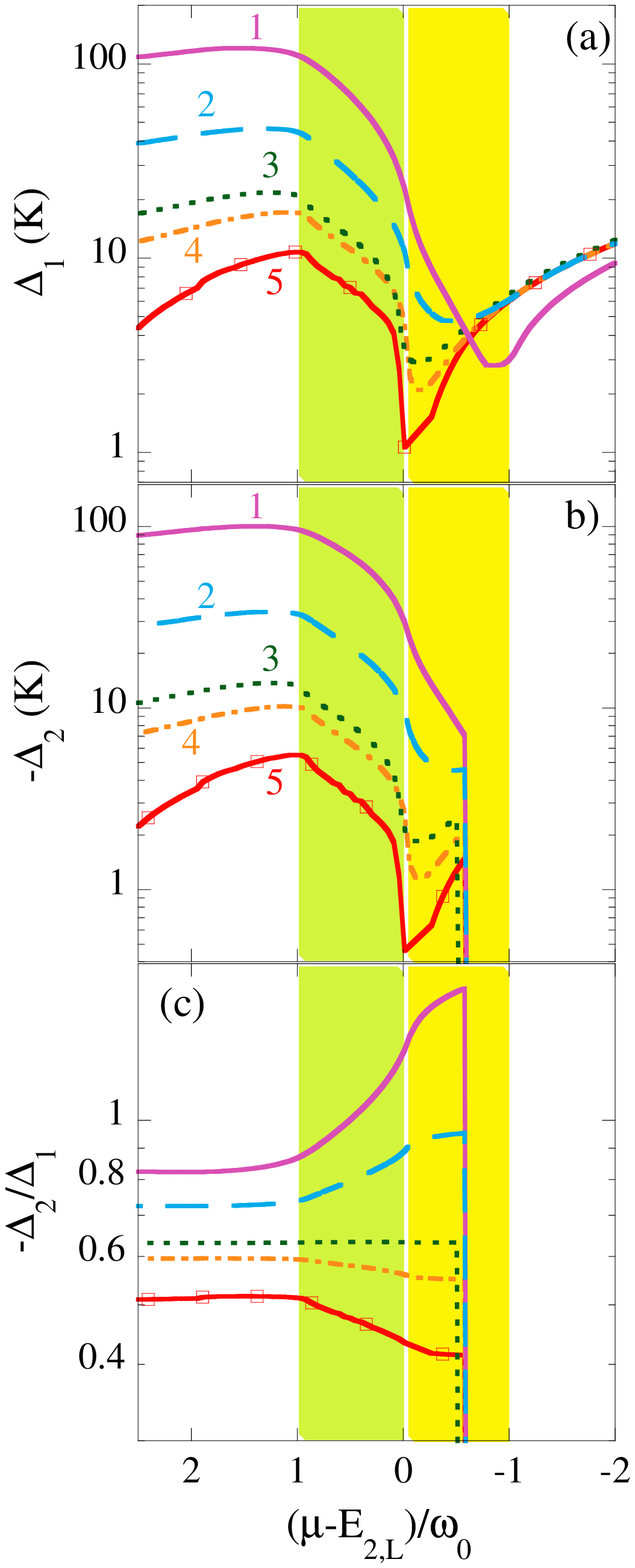}
\caption{(Color online). The evolution of the gap, averaged in the $k_z$ direction, in the
first mini-band [panel (a)] and in the second one [panel (b)] as a function of the Lifshitz
parameter, with $c_{22}/c_{11}=0.45$ and $c_{11}=0.22$. The ratio of the gaps as a function
of the Lifshitz parameter is reported in panel (c). Several interband coupling ratio cases
are reported: curve 1 (solid line) shows the case for $c_{12}/c_{11}=-2.73$, curve 2 (dashed line)
shows the case for $c_{12}/c_{11}=-1.59$, curve 3 (dotted line) shows the case for
$c_{12}/c_{11}=-1.04$, curve 4 (dot-dashed line) shows the case for $c_{12}/c_{11}=-0.91$ and curve
5 (solid line with open squares) shows the case for $c_{12}/c_{11}=-0.68$. The curve 3
(dotted line) shows the critical case of interband pairing
$-c_{12}^{critical}=\sqrt{c_{11}c_{22}}/0.656$ where the gap ratio remains constant going
from the two-gap BCS Fermi case to the mixed Bose-Fermi regime at the band edge.}\label{fig5}
\end{figure}

\begin{figure}[tbp]
\centering
\includegraphics [angle=0,scale=0.45]{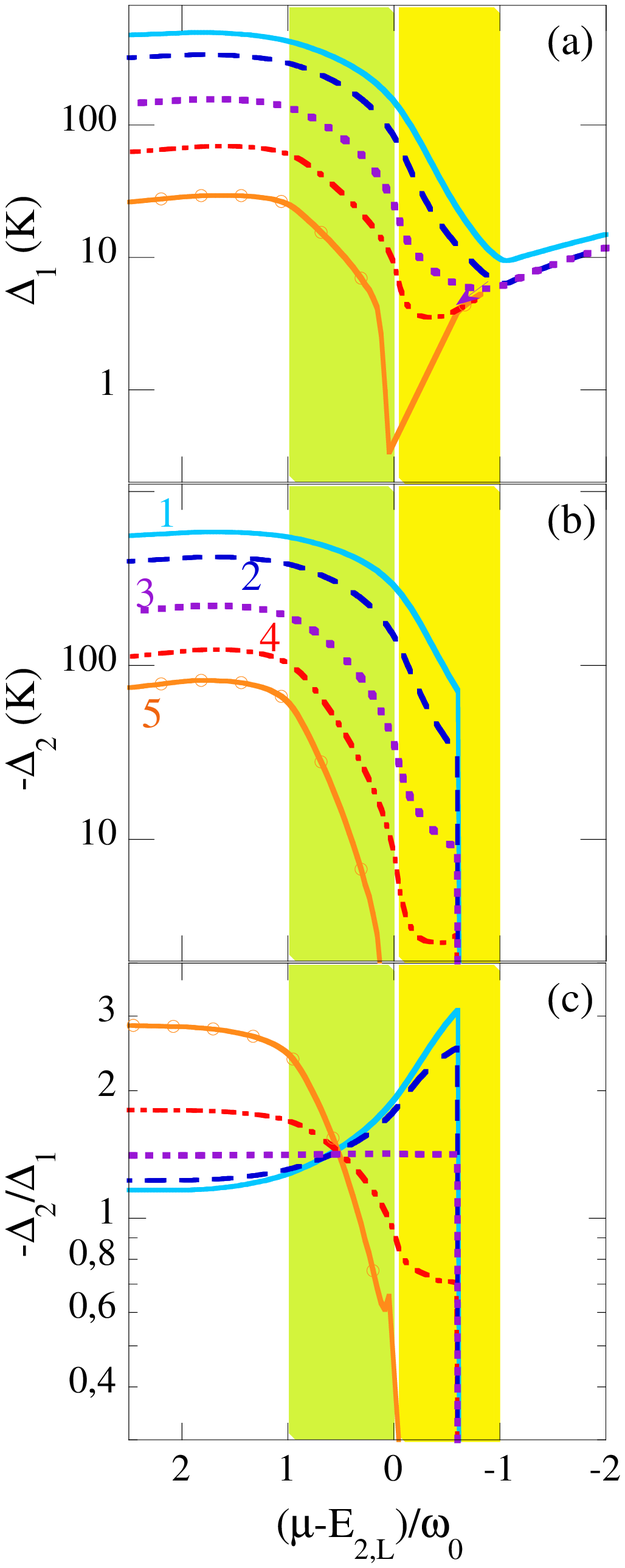}
\caption{(Color online). The evolution of the gap, averaged in the $k_z$ direction, in the
first mini-band [panel (a)] and in the second one [panel (b)] as a function of the Lifshitz
parameter, with $c_{22}/c_{11}=2.17$ and $c_{11}=0.22$. The ratio of the gaps as a function
of the Lifshitz parameter is reported in panel (c). Several interband coupling ratio cases are
reported: curve 1 (solid line) shows the case for $c_{12}/c_{11}=-6.09$, curve 2 (dashed line)
shows the case for $c_{12}/c_{11}=-4.34$, curve 3 (dotted line) shows the case for
$c_{12}/c_{11}=-2.24$, curve 4 (dot-dashed line) shows the case for $c_{12}/c_{11}=-1.08$
and curve 5 (solid line with open squares) shows the case for $c_{12}/c_{11}=-0.43$. The curve 3
(dotted line) shows the critical case of interband pairing
$-c_{12}^{critical}=\sqrt{c_{11}c_{22}}/0.656$ where the gap ratio remains constant from the
two-gap BCS Fermi case to the mixed Bose-Fermi regime at the band edge.}\label{fig6}
\end{figure}

\begin{figure}[tbp]
\centering
\includegraphics [angle=0,scale=0.60]{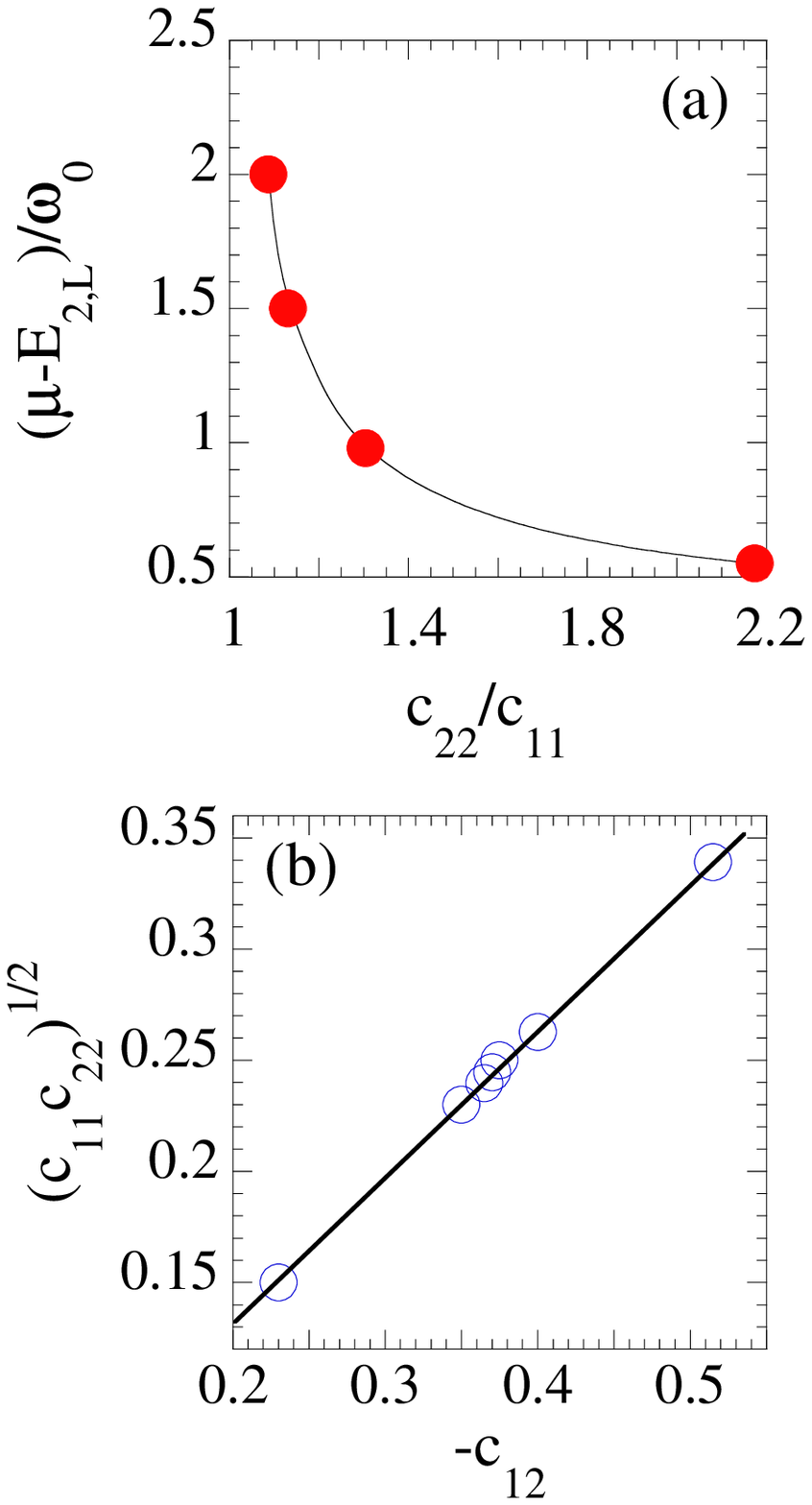}
\caption{(Color online). Panel (a). The variation of the critical chemical potential (solid line with
filled circles) as a function of intraband coupling ratio $c_{22}/c_{11}$. We observe that
for the intraband coupling ratio approaching unity the critical chemical potential goes to infinity.
Panel (b). The variation of the critical interband pairing as a function of $\sqrt{c_{11}c_{22}}$. We
observe a linear fit $\sqrt{c_{11}c_{22}}=-0.656$ $c_{12}^{critical}$ (solid line) of calculated
$c_{12}^{critical}$ points for the cases of $\sqrt{c_{11}c_{22}}$ (open circles) that divide
the plane space in two regions: the upper one, for $\sqrt{c_{11}c_{22}}>-c_{12}^{critical}$, in
weak interband regime, where $\partial(-\Delta_2/\Delta_1)/\partial\mu>0$ and the lower one, for
$\sqrt{c_{11}c_{22}}<-c_{12}^{critical}$, in strong interband regime, where
$\partial(-\Delta_2/\Delta_1)/\partial\mu<0$.}\label{fig7}
\end{figure}

In the standard BCS theory, where $E_F$ is far from the band edges,
the relative variation of $\mu$ going from the normal to the
superconducting state is expected to be negligible. This is not true when
$\mu$ is tuned near the band edge of the second band. In fact,
our calculation yields a significant variation of $\mu$ in
the superconducting phase, as a function of the charge density (Fig. 2b). A
relative variation of $\mu$ going from the normal to the
superconducting phase, as large as $10^{-3}$, is obtained near the band edge and at
the 3D-2D ETT, within a range $4\omega_0$. The variation starts to be large,
as compared with the standard BCS result, in proximity of the crossover
regime below the band edge up to well beyond the 3D-2D ETT.

The Feshbach-like shape resonance regime occurs in correspondence of the large
variation of $\mu$ between the normal and superconducting phases shown in Fig. 2b.
In Fig. 3 we report the distribution of values of the matrix elements
$I_{\mathbf{k},\mathbf{k}'}^{\ell,\ell'}$ in the first two mini-bands, for the case
with mini-band dispersion $\xi=\omega_0$. The intraband distributions of the two bands
show different shapes and widths and have different range of values. The resulting
matrix of coupling constants is obtained exclusively from the eigenfunctions of the
superlattice and is noticeably asymmetric.

The gap parameters in the two bands have a sizable dependence on the
wave-vector $k_z$ as shown in Fig. 4 for the gap in the
first mini-band, taking the average value of the gap over the wave-vectors
as a reference value. The wave-vector dependence of the gaps is induced
by the effective interaction $V_{\mathbf{k},\mathbf{k}'}^{\ell,\ell'}$
entering the BCS equations, which by itself has a non-trivial dependence on
the wave-vectors of the two scattered electrons forming the Cooper pairs,
as discussed above. The largest values of the gaps are found in our calculations
for large transversal wave-vectors, suggesting that the effective pairing
interaction is mostly affected by the superlattice at small distances of the
order of the modulation of the confining periodic potential. Therefore,
all the superconducting properties (coherence length, critical temperature,
gaps to $T_c$ ratios) are expected to be influenced by superlattice effects
via the wave-vector dependence of the gap parameters in different bands.

The variation of the gaps at $T=0$ as function of the chemical potential
in different coupling regimes are plotted in Figs. 5 and 6. In both figures,
the average value of the two gaps taken over the transversal wave-vector $k_z$
is reported. In Fig. 5 we consider the case where the intraband coupling in the
second mini-band is {\em smaller} than the intraband coupling in the first mini-band,
i.e., $c_{22}/c_{11}=0.45$. In particular, we report the gaps in the first
[panel (a)] and second [panel (b)] mini-band as a function of the
chemical potential tuned in the vicinity of the bottom of the second
mini-band. The ratio between the two gaps is reported in the same
figure, panel (c). Several interband couplings are considered, spanning
from $(c_{12}=c_{21}=-0.15)$ to $(c_{12}=c_{21}=-0.60)$.
Interestingly, a pronounced minimum is present for the average value of the gap
at the FS of the first mini-band, $\Delta_1$, when the chemical potential
approaches the bottom of the second mini-band from below. The partial DOS
and the intraband coupling in the first mini-band do not change moving
the chemical potential therefore the presence of a pronounced minimum supports the
existence of antiresonances in the superconducting gaps induced by the interplay
between the wave-vector and band index dependence of the effective interaction
(in which the superposition of single-particle wave-functions gives rise to
interference effects).
The depth of the minimum and its energy position depend on the interband
exchange-like pairing. It generally appears below the band edge, where the DOS
of the second mini-band changes abruptly. It occurs well below the band edge for strong
interband pairing and moves toward the band edge by decreasing the interband pairing.
The details of the antiresonance depend on the interband coupling, but
its presence is a robust feature.

The average value of the superconducting gap at the FS of the second mini-band, $\Delta_2$, appears
below the energy of the band edge where the chemical potential enters in the second
mini-band. It displays a strong interband coupling dependence [see Fig. 5, panel (b)].
The ratio $-\Delta_2/\Delta_1$ reaches a constant energy independent value at a
critical value $c_{12}^{critical}$ of the repulsive exchange-like interband pairing.
The average value of the superconducting gap at the FS of the second mini-band, $\Delta_2$, shows
an antiresonance minimum for $|c_{12}|<|c_{12}^{critical}|$.
The gap ratio $-\Delta_2/\Delta_1$ shows a divergent enhancement moving the chemical potential
from the two-gap BCS regime well above the band edge toward the mixed Bose-Fermi regime near
the edge for $|c_{12}|>|c_{12}^{critical}|$.
The gap ratio $-\Delta_2/\Delta_1$ decreases moving the chemical potential from the two-gap
BCS regime well above the band edge toward the mixed Bose-Fermi regime reaching a minimum at
the band edge and increasing  below the band edge for $|c_{12}|<|c_{12}^{critical}|$, as
shown in Fig. 5, panel (c). When the chemical potential is well inside the second mini-band,
the two gaps recover the standard behavior of a two-band superconductor, with the gap ratio
related to the coupling ratio $c_{22}/c_{11}$. Therefore $-\Delta_2/\Delta_1$ remains smaller
than one in the present case.
The results show clearly that in the crossover regime there is a critical interband pairing
$c_{12}^{critical}$ that separates a first regime of strong interband, where the gap ratio
$-\Delta_2/\Delta_1$ increases with moving the chemical potential toward the band edge
from a second regime of weak interband, where the gap ratio $-\Delta_2/\Delta_1$ decreases
approaching the band edge.

In Fig. 6 we consider the case where the intraband coupling in the
second mini-band is {\em larger} than the intraband coupling in the
first mini-band, i.e., $c_{22}/c_{11}=2.17$.
In particular, the gaps in the first
[panel (a)] and second [panel (b)] mini-band as a function of the
chemical potential tuned at the bottom of the second
mini-band are reported, together with the ratio between the two gaps in
panel (c). The calculations have been carried out for several interband couplings, spanning
from weak $(c_{12}=c_{21}=-0.51)$ to strong interband exchange-like repulsive coupling $(=-1.4)$.
The pronounced antiresonance minimum is clearly present for $\Delta_1$.
The ratio $-\Delta_2/\Delta_1$ reported in panel (c) of Fig. 6 reaches a constant energy independent
value at the critical value $c_{12}^{critical}$ of the repulsive exchange-like interband pairing.

Our results for the case $c_{22}/c_{11}>1$ show that the ratio $-\Delta_2/\Delta_1$ display another
relevant feature: at a critical
position of the chemical potential the ratio between the two gaps becomes universal, i.e.,
independent of the value of the interband coupling. The critical chemical potential in Fig. 6 is above
the bottom of the second mini-band, at a distance about $\frac{1}{2}\omega_0$ from it. This is again
a signature of interference effects in the effective pairing, resulting in a fixed point
between an antiresonance and a resonance in the superconducting gaps.
In Fig. 6  we observe that well above the band edge, in the standard two-gap BCS regime, the gap ratio
$-\Delta_2/\Delta_1$ is larger than one, when such is the ratio between the intraband couplings.
Increasing the interband pairing we observe that there is a value of the chemical potential
at which there is a reversal of the ratio $-\Delta_2/\Delta_1$.

The critical point of the chemical potential where the gap ratio is independent of the interband
pairing for $c_{22}/c_{11}>1$ depends on the ratio between the intraband attractive couplings
in the two bands. The variation of the critical chemical potential, present only for the case
$c_{22}>c_{11}$, depends on the ratio between the intraband couplings $c_{22}/c_{11}>1$, as it is
shown in Fig. 7 [panel (a), dashed blue line]. The critical chemical potential goes to infinity
when the couplings are equal and decreases toward the band edge with increasing the ratio between
the coupling in the second band on the coupling in the first band.
The critical interband pairing $c_{12}^{critical}$ increases linearly with increasing
$\sqrt{c_{11}c_{22}}$, as it is shown in Fig. 7 [panel (b), solid red line].
The data can be fitted with the linear relation $\sqrt{c_{11}c_{22}}=-0.656\,c_{12}^{critical}$,
as it is shown in Fig. 7 [panel (b)].

\section{The critical temperature}

The superconducting critical temperature $T_c$ is calculated by iteratively solving the linearized
gap equation
\begin{equation}
\Delta_{\ell,\mathbf{k}}=
-\frac{1}{2M}\sum_{\ell',\mathbf{k}'}\mathcal{V}_{\mathbf{k},\mathbf{k}'}^{\ell,\ell'}
\ \frac{\tanh(\frac{\xi_{\ell',\mathbf{k}'}}{2T_c})}{\xi_{\ell',\mathbf{k}'}}\Delta_{\ell',\mathbf{k}'},
\label{7}
\end{equation}
until the non-trivial solution (having eigenvalue 1) is reached with increasing temperature.

\begin{figure}[tbp]
\centering
\includegraphics [angle=0,scale=0.62]{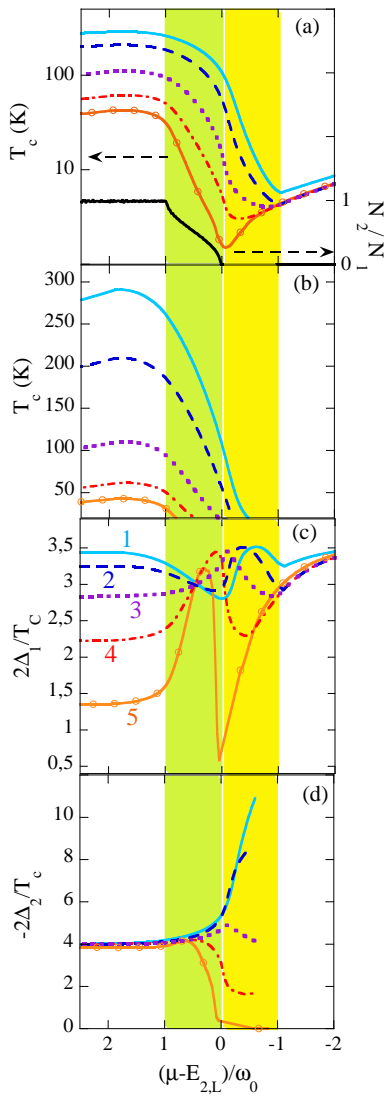}
\caption{(Color online). The case of strong coupling in the second band (the
so-called diboride case) with
$c_{22}/c_{11}=2.17$ and $c_{11}=0.22$. The variable ratio
$u=N_2/N_1$ (solid black line) is shown in panel (a). The critical temperature
$T_c$ [panels (a) and (b)], the ratio $2\Delta_1/T_c$ [panel (c)], the ratio $-2\Delta_2/T_c$
[panel (d)],
as functions of the reduced Lifshitz parameter $(\mu-E_{2,L})/\omega_0$.
Several values of the interband coupling ratio are reported: curve 1 (solid line) shows the case for
$c_{12}/c_{11}=-6.09$, curve 2 (dashed line) shows the case for $c_{12}/c_{11}=-4.34$, curve 3
(dotted line) shows the case for$c_{12}/c_{11}=-2.24$, curve 4 (dot-dashed line) shows the case
for $c_{12}/c_{11}=-1.08$ and curve 5 (solid line with open squares) shows the case for
$c_{12}/c_{11}=-0.43$.
For the largest value of $c_{12}$ considered here, the critical temperature reaches 300K, in
the range 1$<(\mu-E_{2,L})/\omega_0<$2. At the antiresonance for
$(\mu-E_{2,L})/\omega_0=0$, $T_c$ and $2\Delta_1/T_c$ reach the minimum value for weak
interband interaction. On the contrary, for large interband interaction,
$2\Delta_1/T_c$ and $T_c$ reach the minimum value at
$(\mu-E_{2,L})/\omega_0=-1$, where $T_c$ reaches room temperature for the
chosen set of parameters.}
\label{fig8}
\end{figure}

Below, we present numerical results for the solution of the self-consistent
linearized BCS equations which determines the value of the critical
temperature and of the chemical potential.
We analyze the behavior of the ratios $2\Delta_1/T_c$ and $2\Delta_2/T_c$,
where, according to our notation, $\Delta_1$ and $\Delta_2$ are the average values of
the wave-vector dependent gaps on the corresponding branches of the FS at $T=0$.
Note that the gaps to $T_c$ ratios are physical quantities
which are very suitable for quantitative comparison with experiments.
Indeed, gaps at $T=0$ (or temperature below $0.4\,T_c$ in real experimental
conditions), and $T_c$ can be extracted in a single experimental run
when measuring the temperature dependence of the superconducting gaps
in multiband superconductors.

We focus here on the case characterized by a coupling in
the second FS larger than in the first FS. As an example of this regime, we calculate
the DOS for the second mini-band, $T_c$,
$2\Delta_1/T_c$ and $-2\Delta_2/T_c$ as functions of the reduced Lifshitz parameter
$(\mu-E_{2,L})/\omega_0$, with intraband
coupling terms fixed to $c_{11}=0.23$, $c_{22}/c_{11}=2.17$, for several values of the
interband coupling ratio, as shown in the panels (a-d) of Fig. 8.

The minima of $T_c$, due to the \emph{shape antiresonances}, are all below the band edge energy
of the second mini-band [panel (a)] while the maximum value of $T_c$, due to the \emph{shape resonance},
is reached near the 3D-2D ETT of the second mini-band, on its 2D side [panel (b)].
We observe that the \emph{shape resonances} in the gap to $T_c$ ratios show clear evidence for the
typical asymmetric line shape of Fano-Feshbach quantum resonances driven by configuration interaction
between different scattering channels. The antiresonance due to negative interference effects appears
in the range between $(\mu-E_{2,L})/\omega_0=-1$ and $(\mu-E_{2,L})/\omega_0=0$, where $T_c$ has a
minimum in panel (a) followed by the $T_c$ resonance maximum in the range $(\mu-E_{2,L})/\omega_0=1.5$.

For strong exchange-like interband pairing, $c_{12}>c_{12}^{critical}$, we find that the ratio
$2\Delta_1/T_c$ shows two minima and the ratio $2\Delta_2/T_c$ shows an increasing divergence in
the mixed Bose-Fermi regime. Decreasing the interband pairing, for $c_{12}<c_{12}^{critical}$, the
minimum due to the negative interference effect in $2\Delta_1/T_c$ at $(\mu-E_{edge})/\omega_0=-1$,
associated with the Feshbach-like \emph{shape resonance} at the band edge, moves toward
$(\mu-E_{edge})/\omega_0=0$. The ratio $2\Delta_1/T_c$ shows a maximum at $(\mu-E_{edge})/\omega_0=0$
[dotted line in panel (c) of Fig. 8] for the critical value of the interband pairing,
$c_{12}^{critical}$.

\section{The case of the superlattice of atomic boron layers: doped $MgB_2$}

\begin{figure}[tbp]
\centering
\includegraphics [angle=0,scale=0.55]{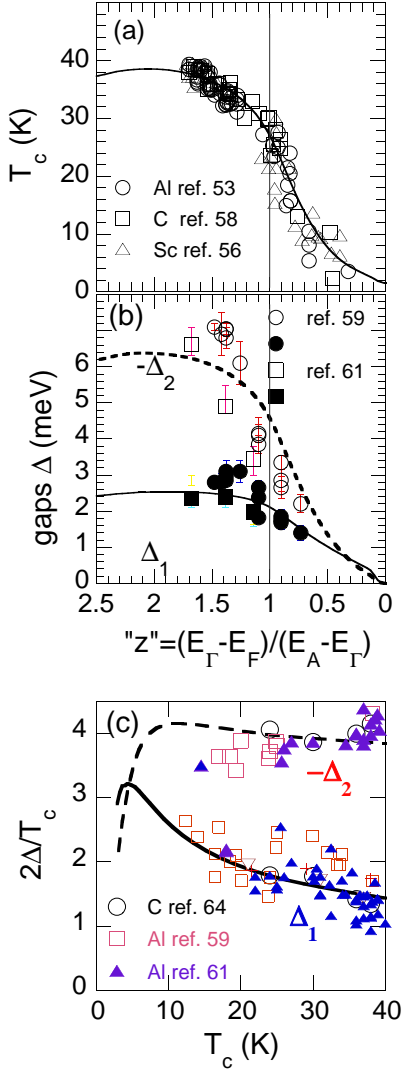}
\caption{(Color online).
Panel (a). The superconducting critical temperature in Al for Mg substitution $(Mg_{1-x}Al_x)B_2$,(open circles)\cite{15x}, in Sc for Mg substitution $(Mg_{1-x}Sc_x)B_2$,(open triangles)\cite{15y} and for the C for B substitution in the $Mg(C_{x}B_{1-x})_2$ system (open squares)\cite{carbon} as function of the Lifshitz parameter $"z"=(E_{\Gamma}-E_F)/(E_A-E_{\Gamma})$ for diborides.\cite{15x,15y,carbon}
Panel (b). The gaps $\Delta_{1}=\Delta_{\sigma}$ (open symbols) and $\Delta_{2}=\Delta_{\pi}$ (filled symbols) in Al for Mg substitution $(Mg_{1-x}Al_x)B_2$,\cite{15xDaghero}  and for the C for B substitution in the $Mg(C_xB_{1-x})_2$ system\cite{15zShin}  as function of the Lifshitz parameter $"z"=(E_{\Gamma}-E_F)/(E_A-E_{\Gamma})$ calculated for aluminum\cite{15x,15xx} and carbon\cite{15z} substitutions based on theoretical calculations and structural data.
Panel (c).  The BCS ratios curves $2\Delta_1/T_c$ (solid line) and $-2\Delta_2/T_c$ (dashed line), as functions of $T_c$, extracted from the experiments measuring in the same run the two gaps well below the critical temperature and $T_c$.\cite{15xDaghero,15xSzabo1,15xSzabo2,15xKlie,15xPutti}Carbon doped samples,\cite{15zShin} (open circles). Al doped samples from Daghero et al.\cite{15xDaghero} (open squares), and from Samuely et al.\cite{15xSzabo2} (triangles).
The experimental data are compared with the calculated curves in the case $c_{22}/c_{11}= 2.17$, $c_{12}/c_{11}=-0.43$, where $-c_{12}=0.10$ is lower than the value $(-c_{12}^{critical})=0.52$, where the critical temperature, and the gaps are calculated as a function of $(\mu-E_{2,L})/\xi_0$.
}
\label{fig9}
\end{figure}

The present shape resonance superconductivity scenario in a multiband superconductor tuning the chemical
potential near a band edge can be tested in the particular experimental case of doped magnesium diboride.
Doped MgB$_2$ is a practical realization of a heterostructure at atomic limit\cite{patent} being made of
first units, superconducting atomic boron layers (similar to graphene layers), intercalated with
second units (playing the role of spacers), hexagonal Mg layers, forming a superlattice of quantum
wells with a period of 0.35 nm.\cite{15a1,15a} The electronic structure near the Fermi level resulting
from the material architecture is made of a first large Fermi surface ($\pi$-band) and second small
Fermi surface ($\sigma$-band) with different symmetry and spatial locations. The DOS at the Fermi energy
in MgB$_2$ is 0.12 states/(eV$\cdot$atom$\cdot$spin), where about one half (exactly $44\%$) of this
value comes from the $\sigma$-bonds and the other half (exactly 56\%) comes
from the $\pi$-bonds. The chemical potential $E_F$ in MgB$_2$ is separated by about 700 meV from the
band edge energy at the top $E_{\Gamma}$ of the $\sigma$-band and it is separated by about 300 meV
from the energy of the 3D-2D ETT at the $E_A$ point in the band structure.

The chemical potential can be tuned near the band edge of the sigma band (at the critical point
$E_{edge}=E_{\Gamma}$ in the band structure) and across the 3D-2D ETT
(at the critical point $E_{3D-2D}=E_{A}$ in the band structure) by chemical substitution of
Al\cite{15a,15x,15xx,15xxx} and Sc\cite{15y} for Mg or C for B.\cite{15z,carbon}  The different
superconducting gaps in both bands and the critical temperature have been  measured as a function of
the chemical potential across the 3D-2D ETT by several groups in Al doped
samples,\cite{15xDaghero,15xSzabo1,15xSzabo2,15xKlie,15xPutti} and in carbon doped samples.\cite{15zShin}
The physical properties around the 3D-2D ETT are plotted here for the case of doped MgB$_2$ as a function
of the reduced Lifshitz parameter $"z"=(E_{\Gamma}-E_F)/(E_A-E_{\Gamma})$, where
$\xi_0$=$E_A-E_{\Gamma}$ is the energy dispersion in the c-axis direction due to electron hopping
between the boron layers (400 meV in MgB$_2$ and it changes with chemical substitution x). The influence
of the proximity to an ETT\cite{olij} on the anomalous electronic and lattice properties of MgB$_2$ is
shown by the anomalous pressure dependence\cite{goncharov} and Raman lineshape\cite{simonelli} of the
$E_2g$ phonon mode.

The multiband superconducting phase of MgB$_2$ in the clean limit has been studied theoretically in
detail\cite{15choi,15choi2,21,15ummarino} and its properties have been extracted from experimental
data.\cite{10,15zShin} The ratio between the intraband electron-phonon (p-h) coupling in the
$\sigma$-band, $c_{22}$, and in the $\pi$-band, $c_{11}$, is in the range $2<c_{22}/c_{11}<2.6$
and the ratios between the interband coupling constant and the intraband coupling constants are
$0.15<c_{12}/c_{22}<0.3$ and $0.4<c_{12}/c_{11}<1$. The system is therefore in the regime of strong
coupling in the second band and weak interband coupling discussed above. The interband coupling is
smaller than the critical interband coupling $-c_{12}^{critical}=\sqrt{c_{11}c_{22}}/0.656$: from
the published data we find $0.2<c_{12}/(-c_{12}^{critical})<0.5$. Therefore the shape resonance regime
for the doped MgB$_2$ system is well described by the typical case in Fig. 6 shown as the curve 5 where
$c_{22}/c_{11}= 2.17$, $c_{12}/c_{11}=-0.43$ and $c_{12}/(-c_{12}^{critical})=0.19$.
For diborides there is a lack of a focused experimental test for the so called
$s\pm$ pairing, driven by the repulsive Coulomb interaction in the two band superconductivity with negative interband coupling. This will lead to condensate wavefunctions with opposite signs as it is proposed here. Therefore we hope that an experiment like for pnictides\cite{16} will be run in the near future. However the critical temperature and the absolute values of the gaps calculated here does not depend in the sign of the interband coupling.
The critical temperature (panel a) superconducting gaps (panel b) in doped MgB$_2$ are shown in Fig. 9 as
a function of the Lifshitz parameter "z". The Lifshitz parameter "z" has been calculated at each chemical
doping value x in the case of the Al substitution for Mg $(Mg_{1-x}Al_x)B_2$,\cite{15x,15xx,15xxx} for
the case of Sc substitution for Mg in the $(Mg_{1-x}Sc_x)B_2$ system\cite{15y} and for the C for B
substitution in the $Mg(C_{x}B_{1-x})_2$ system\cite{carbon,15z} based on the measured variation of the
lattice parameters and the charge density using band structure calculations.\cite{15x,15xx,15xxx,15z}
Fig. 9a shows the universal scaling of the critical temperature $T_c$, as a function of the
Lifshitz parameter "z" while the doping mechanism is quite different for the aluminum,\cite{15x,15xx}
scandium\cite{15y} and carbon\cite{carbon,15z} chemical substitutions in diborides. It is clear that
the critical temperature curves collapse on the same curve for all cases when they are plotted as
a function of "z". The experimental data are plotted in the same figure with the theoretical
calculations for the generic $T_c$ and gaps behavior for the case 5 in Fig. 6 and it is possible
to appreciate the good agreement between theory and experimental data. In Fig. 9c, the BCS gap
ratios $2\Delta_1/T_c$ and $-2\Delta_2/T_c$ of aluminum doped and carbon doped diborides are plotted
as functions of $T_c$ and are compared with the same theoretical case discussed above.
The agreement between our theoretical calculations
and the available experimental data is satisfactory for the dependence of $T_c$, the
gaps, and the BCS ratio in a wide range of aluminum doping of magnesium diborides. The difference between
the experimental data and the present calculations is related with the fact the
intraband coupling in the second band is likely to depend on the chemical
potential,\cite{simonelli} whereas we assume it to be constant, to reduce the
number of parameters. Further work is under progress for a quantitative investigation of shape resonances
in doped diborides.
These results show clearly that superconductivity in doped diboride superlattices remain in the clean limit
for interband pairing although the large number density of impurity centers in doped samples and the
structural phase separation.\cite{simonelli} Indeed, it has been pointed out that the shape
resonance mechanism can be favored by phase separation and especially for fractal distributions
in superstripes.\cite{35,18,15g}

\section{Conclusions}

The theoretical analysis discussed in this paper was motivated by the
crucial observation that interband pairing may yield a sizable enhancement
of the critical temperature in multiband materials, displaying superconductivity
in the clean limit, possibly leading to room temperature superconductivity,\cite{patent}
and is the first step toward the understanding of the role of \emph{shape resonances}
in the superconducting gaps in multilayer systems. Although limited to the BCS
mean-field approximation, our approach captures the main ingredients of the physics
of these systems. In particular, once the relevant details of the wave
functions and of the band dispersions of the multilayer structure are introduced in
the coupled BCS-like equations for the gaps in the various bands and for the conservation
of the electron density, the resonance physics can be investigated in the very same spirit
of the BCS-BEC crossover in fermion gases, where a similar approach proved to be
predictive.

In real materials, the shape resonances can be controlled with material science techniques,
tuning the pairing strength and tailoring the thickness of the spacer layers, to adjust the
electron hopping
between superconducting layers. In our theoretical approach, the crucial parameter to take into
account the effects of hopping variations on the characteristics of the shape resonance, is
the dispersion $\xi$. Resonance effects are emphasized when $\xi$ is on the order of the
energy cutoff $\omega_0$ of the effective pairing interaction leading to Cooper pairing. In this
paper, we devoted our analysis to the case $\xi=\omega_0$, but of course $\xi$ can be varied
around the optimal amplification value, and an analysis of the dependence of resonance effects
on the value of $\xi$ will be the object of future investigation.

The main outcomes of our theoretical scenario for superconducting multilayers
or multigap superconductors have been tested here on aluminum-doped magnesium diboride
superconductors. The quantitative comparison between our theoretical results and available
experimental data for this material is satisfactory and permits to identify, e.g., the relevant
range of values for the pairing coupling constants.

Our approach can be easily extended to pursue a preliminary investigation of the
superconducting properties of other multiband system where the chemical potential is
tuned near a  ETT  in one of the bands. For instance, it could be applied to
Fe-based 1111, 122 and 11 pnictides.

On the basis of our analysis, we may start to indicate
a possible roadmap for the discovery of novel HTS.\cite{patent}
The conditions for higher temperature superconductivity, or even room temperature
superconductivity, might be met in graphene bilayers and graphane,\cite{19,20}
where the exchange-like interband interaction could be even stronger than the intraband
Cooper pairing, as also predicted for superlattices of carbon nanotubes.\cite{18}

\begin{acknowledgments}
We thank Rocchina Caivano, and Valeria D'Andrea for help in the early stage of this work. We are
grateful to Arkady A. Shanenko and I. Eremin for useful discussions.
\end{acknowledgments}


\begin{thebibliography}{9}


\bibitem{Leggett} A. J. Leggett, in Modern Trends in the Theory
of Condensed Matter, edited by by A. Pekalski and R. Przystawa, Lecture Notes in Physics Vol. 115
(Springer-Verlag, Berlin, 1980), p. 13.

\bibitem{Perali04}
A. Perali, P. Pieri, and G.C. Strinati, Phys. Rev. Lett. 93, 100404 (2004), and references therein.

\bibitem{PeraliPRA2003} A. Perali, P. Pieri, and G.C. Strinati, Phys. Rev. A 68, 031601 (2003).

\bibitem{GiorginiRMP2008} S. Giorgini, L.P. Pitaevskii, and S. Stringari,
Rev. Mod. Phys. 80, 1215 (2008), and reference therein.

\bibitem{30}
J. M. Blatt and C. J. Thompson, Physical Review Letters 10, 332 (1963), URL http://dx.doi.org/10.1103/PhysRevLett.10.332.

\bibitem{601}
A. A. Shanenko, M. D. Croitoru, and F. M. Peeters, EPL (Europhysics Letters) pp. 498+ (2006), URL http://dx.doi.org/10.1209/epl/i2006-10274-6.

\bibitem{31}
C. Thompson, Journal of Physics and Chemistry of Solids 26, 1053 (1965), ISSN 00223697, URL http://dx.doi.org/10.1016/0022-3697(65)90194-0.

\bibitem{34}
A. Perali, A. Bianconi, A. Lanzara, and N. L. Saini, Solid State Communications 100, 181 (1996), ISSN 00381098, URL http://dx.doi.org/10.1016/0038-1098(96)00373-0

\bibitem{32}
A. A. Shanenko and M. D. Croitoru, Physical Review B 73, 012510+ (2006), URL http://dx.doi.org/10.1103/PhysRevB.73.012510.


\bibitem{kresin}
V. Z. Kresin and Y. N. Ovchinnikov, Physical Review B 74, 024514+ (2006), URL http://dx.doi.org/10.1103/PhysRevB.74.024514.

\bibitem{guo}
Y. Guo, Y.-F. Zhang, X.-Y. Bao, T.-Z. Han, Z. Tang, L.-X. Zhang, W.-G. Zhu, E. G. Wang, Q. Niu, Z. Q. Qiu, et al., Science 306, 1915 (2004), URL http://dx.doi.org/10.1126/science.1105130.

\bibitem{bose1}
S. Bose, P. Raychaudhuri, R. Banerjee, P. Vasa, and P. Ayyub, Physical Review Letters 95, 147003+ (2005), URL http://dx.doi.org/10.1103/PhysRevLett.95.147003.

\bibitem{Zhang}
Y. F. Zhang, J. F. Jia, T. Z. Han, Z. Tang, Q. T. Shen, Y. Guo, Z. Q. Qiu, and Q. K. Xue, Physical Review Letters 95, 096802+ (2005), URL http://dx.doi.org/10.1103/PhysRevLett.95.096802.

\bibitem{brun}
C. Brun, I. P. Hong, F. Patthey, Yu, R. Heid, P. M. Echenique, K. P. Bohnen, E. V. Chulkov, and W. D. Schneider, Physical Review Letters 102, 207002+ (2009), URL http://dx.doi.org/10.1103/PhysRevLett.102.207002.

\bibitem{bose2}
S. Bose, A. M. Garcia-Garcia, M. M. Ugeda, J. D. Urbina, C. H. Michaelis, I. Brihuega, and K. Kern, Nature Materials advance online publication (2010), ISSN 1476-1122, URL http://dx.doi.org/10.1038/nmat2768.

\bibitem{patent}
A. Bianconi, "Process of increasing the critical temperature Tc of a bulk superconductor by making metal heterostructures at the atomic limit" US Patent No. 6,265,019 (2001).

\bibitem{france}
A. Bianconi and M. Missori, Journal de Physique I 4, 361 (1994), ISSN 1155-4304, URL http:
//dx.doi.org/10.1051/jp1:1994100.

\bibitem{SSC}
A. Bianconi, Solid State Communications 89, 933 (1994), ISSN 00381098, URL http://dx.doi.org/10.1016/0038-1098(94)90354-9.

\bibitem{15d}
A. Bianconi, Journal of Superconductivity and Novel Magnetism 18, 25 (2005), ISSN 1557-1939, URL http://dx.doi.org/10.1007/s10948-005-0047-5.

\bibitem{15f}
A. Bianconi and M. Filippi in ÓSymmetry and Heterogeneity in high temperature superconductorsÓ edited by A. Bianconi (Spinger Dordrecht 2006)  Nato Science Series II Mathematics, Physics and Chemistry 214, 21-53 (2006). ISBN-13: 978-1402039881.

\bibitem{35}
A. Bianconi, A. Valletta, A. Perali, and N. L. Saini, Physica C: Superconductivity 296, 269 (1998), ISSN 09214534, URL http://dx.doi.org/10.1016/S0921-4534(97)01825-X.

\bibitem{18}
A. Bianconi, Physica Status Solidi (a) 203, 2950 (2006), ISSN 18626300, URL http://dx.doi.org/10.1002/pssa.200567003.

\bibitem{15g}
Rocchina Caivano, Michela Fratini, Nicola Poccia, Alessandro Ricci, Alessandro Puri, Zhi-An Ren, Xiao-Li Dong, Jie Yang, Wei Lu, Zhong-Xian Zhao, Luisa Barba and Antonio Bianconi Supercond. Sci. Technol. 22, 014004 (2009) doi: 10.1088/0953-2048/22/1/014004 ; http://dx.doi.org/10.1007/s10948-008-0433-x.

\bibitem{SS}
A. Bianconi, N. L. Saini, S. Agrestini, D. Di Castro, and G. Bianconi, International Journal of Modern Physics B (IJMPB) 14, 3342 (2000).

\bibitem{35a}
A. Bianconi, D. Di Castro, N. L. Saini, and G. Bianconi, in Phase Transitions and Self-Organization in Electronic and Molecular Networks, edited by M. F. Thorpe and J. C. Phillips (Kluwer Academic Publishers, Boston, 2002), Fundamental Materials Research, chap. 24, pp. 375-388, ISBN 0-306-46568-X, URL $http://dx.doi.org/10.1007/0-306-47113-2_24$.

\bibitem{nature}
M. Fratini, N. Poccia, A. Ricci, G. Campi, M. Burghammer, G. Aeppli, and A. Bianconi, Nature 466, 841 (2010), ISSN 0028-0836, URL http://dx.doi.org/10.1038/nature09260.

\bibitem{15a}
A. Bianconi, D. Di Castro, S. Agrestini, G. Campi, N. L. Saini, A. Saccone, S. De Negri and M. Giovannini, J. Phys. : Condens. Matter 13, 7383-7390 (2001).

\bibitem{graf1}
T. E. Weller, M. Ellerby, S. S. Saxena, R. P. Smith, and N. T. Skipper, Nature Physics 1, 39 (2005), ISSN 1745-2473, URL http://dx.doi.org/10.1038/nphys0010.

\bibitem{graf2}
G. Csanyi, P. B. Littlewood, A. H. Nevidomskyy, C. J. Pickard, and B. D. Simons, Nature Physics 1, 42 (2005), ISSN 1745-2473, URL http://dx.doi.org/10.1038/nphys119.

\bibitem{graf3}
K. Sugawara, T. Sato, and T. Takahashi, Nature Physics 5, 40 (2008), ISSN 1745-2473, URL http://dx.doi.org/10.1038/nphys1128.

\bibitem{15h}
A. Ricci, B. Joseph, N. Poccia, W. Xu, D. Chen, W. S. Chu, Z. Y. Wu, A. Marcelli, N. L. Saini, and A. Bianconi, Superconductor Science and Technology 23, 052003+ (2010), ISSN 0953-2048, URL http://dx.doi.org/10.1088/0953-2048/23/5/052003.

\bibitem{15h1}
Johnpierre Paglione  and  Richard L. Greene "High-temperature superconductivity in iron-based materials" Nature Physics 6, 645 - 658 (2010). doi:10.1038/nphys1759

\bibitem{54}
I. M. Lifshitz, Zh. Eksp. Teor. Fiz. 38, 1569 (1960) [Sov. Phys. JETP 11, 1130 (I960)].

\bibitem{56}
S. Agrestini, N. L. Saini, G. Bianconi, and A. Bianconi, Journal of Physics A: Mathematical and General 36, 9133 (2003), URL http://dx.doi.org/10.1088/0305-4470/36/35/302.

\bibitem{57}
M. Fratini, N. Poccia, and A. Bianconi, Journal of Physics: Conference Series 108, 012036+ (2008), ISSN 1742-6596, URL http://dx.doi.org/10.1088/1742-6596/108/1/012036.
\bibitem{58}
N. Poccia, A. Ricci, and A. Bianconi, Advances in Condensed Matter Physics 2010, 261849 (2010), ISSN 1687-8108, URL http://dx.doi.org/10.1155/2010/261849.

\bibitem{86}
K. Ueno, S. Nakamura, H. Shimotani, A. Ohtomo, N. Kimura, T. Nojima, H. Aoki, Y. Iwasa, and M. Kawasaki, Nature Materials 7, 855 (2008), ISSN 1476-1122, URL http://dx.doi.org/10.1038/nmat2298.
\bibitem{87}
J. T. Ye, S. Inoue, K. Kobayashi, Y. Kasahara, H. T. Yuan, H. Shimotani, and Y. Iwasa, Nature Materials 9, 125 (2009), ISSN 1476-1122, URL http://dx.doi.org/10.1038/nmat2587.
\bibitem{88}
K. Prassides, Nature Materials 9, 96 (2010), ISSN 1476-1122, URL http://dx.doi.org/10.1038/nmat2616.

\bibitem{topological}
X.-L. Qi and S.-C. Zhang, preprint arXiv:1008.2026 (2010), URL http://arxiv.org/abs/1008.2026.

\bibitem{eagles}
E. M. Eagles Phys. Rev. 186, 456 (1969).

\bibitem{vittorini}
Alessandra Vittorini-Orgeas and Antonio Bianconi J Supercond Nov Magn 52, 215-221 (2009) DOI 10.1007/s10948-008-0433-x

\bibitem{cu} S. Caprara, unpublished.


\bibitem{PeraliPRB1998} A. Perali, C. Grimaldi, L. Pietronero, Phys. Rev. B
58, 5736 (1998), and references therein.

\bibitem{22}
N. Kristoffel, P. Konsin, and T. Ord. Rivista Nuovo Cim., 17, 1 (1994).
\bibitem{23}
A. Bussmann-Holder in, \emph{Superconductivity in complex systems}, Structure and bonding, edited by K. A. Muller and A. Bussmann-Holder, vol. 114 (Springer, 2005), ISBN 978, URL http://www.worldcat.org/isbn/978.
\bibitem{21}
E. J. Nicol and J. P. Carbotte, Physical Review B 71, 054501+ (2005), URL http://dx.doi.org/10.1103/PhysRevB.71.054501.

\bibitem{Perali2gap} A. Perali, C. Castellani, C. Di Castro, M. Grilli, E. Piegari, and A. A. Varlamov, Phys. Rev. B 62, R9295 (2000).

\bibitem{16}
C. T. Chen, C. C. Tsuei, M. B. Ketchen, Z. A. Ren, and Z. X. Zhao, Nature Physics 6, 260 (2010), ISSN 1745-2473, URL http://dx.doi.org/10.1038/nphys1531.

\bibitem{16a}
K. Kuroki, S. Onari, R. Arita, H. Usui, Y. Tanaka, H. Kontani, and H. Aoki, Physical Review Letters 101, 087004+ (2008), URL http://dx.doi.org/10.1103/PhysRevLett.101.087004.

\bibitem{Entel}
P. Entel and D. Rainer J. Low Temperature Physics 23, 511 (1976)

\bibitem{15a1}
A. Bianconi, D. Di Castro, S. Agrestini, G. Campi, N. L. Saini "Tc Amplification by Shape Resonance in MgB2" APS, Seattle, March Meeting (2001), MgB2 session, talk 72.

\bibitem{15x}
A. Bianconi, S. Agrestini, D. Di Castro, G. Campi, G. Zangari, N. L. Saini, A. Saccone, S. De Negri, M. Giovannini, G. Profeta, et al., Physical Review B 65, 174515+ (2002), URL http://dx.doi.org/10.1103/PhysRevB.65.174515.

\bibitem{15xx}
O. de la Pena, A. Aguayo, and R. de Coss Physical Review B 66, 012511 (2002)

\bibitem{15xxx}
R. F. Klie, J. C. Zheng, Y. Zhu, A. J. Zambano, and L. D. Cooley, Physical Review B 73, 014513+ (2006), URL http://dx.doi.org/10.1103/PhysRevB.73.014513.

\bibitem{15y}
S. Agrestini, C. Metallo, M. Filippi, L. Simonelli, G. Campi, C. Sanipoli, E. Liarokapis, S. De Negri, M. Giovannini, A. Saccone, et al., Physical Review B 70, 134514+ (2004), cond-mat/0408095, URL http://dx.doi.org/10.1103/PhysRevB.70.134514.


\bibitem{15z}
O. De la Pena-Seaman, R. de Coss, R. Heid, and K.P. Bohnen Physical Review B 79, 134523 (2009).

\bibitem{carbon}
A. Bharathi, S. J. Balaselvi, S. Kalavathi, G. L. N. Reddy, V. S. Sastry, Y. H. Hariharan, and T. S. Radhakrishnan, Physica C: Superconductivity 370, 211 (2002), ISSN 09214534, URL http://dx.doi.org/10.1016/S0921-4534(02)01139-5. ;  S. Lee, T. Masuia, A. Yamamoto, H. Uchiyama, and S. Tajima, Physica C: Superconductivity 397, 7 (2003), ISSN 09214534, URL http://dx.doi.org/10.1016/S0921-4534(03)01296-6.

\bibitem{15xDaghero}
D. Daghero, R. S. Gonnelli, A. Calzolari, G. A. Ummarino, V. Dellarocca, V. A. Stepanov, N. Zhigadlo, S. M. Kazakov, and J. Karpinski, Physica Status Solidi (c) 2, 1656 (2005), ISSN 1610-1634, URL http://dx.doi.org/10.1002/pssc.200460807. The superconducting gaps of C-substituted and Al-substituted MgB2 single crystals by point-contact spectroscopy

\bibitem{15xSzabo1}
P. Samuely, P. Szabo«, P. C. Canfield, and S. L. Bud'ko, Phys. Rev. Lett. 95, 099701 (2005).

\bibitem{15xSzabo2}
P. Samuely, Z. Holanova, P. Szabo, H. T. Wilke, S. L. Bud'ko, and P. C. Canfield, Physica Status Solidi (c) 2, 1743 (2005), ISSN 1610-1634, URL http://dx.doi.org/10.1002/pssc.200460823.

\bibitem{15xKlie}
L. D. Cooley, A. J. Zambano, A. R. Moodenbaugh, R. F. Klie, Jin-Cheng Zheng, and Yimei Zhu, Phys. Rev. Lett. 95, 267002 (2005).

\bibitem{15xPutti}
M. Putti, C. Ferdeghini, M. Monni, I. Pallecchi, C. Tarantini, P. Manfrinetti, A. Palenzona, D. Daghero, R. S. Gonnelli, and V. A. Stepanov, Physical Review B 71, 144505+ (2005), URL http://dx.doi.org/10.1103/PhysRevB.71.144505.

\bibitem{15zShin}
S. Tsuda, T. Yokoya, T. Kiss, T. Shimojima, S. Shin, T. Togashi, S. Watanabe, C. Zhang, C. T. Chen, S. Lee, et al., Physical Review B 72, 064527+ (2005), URL http://dx.doi.org/10.1103/PhysRevB.72.064527.

\bibitem{olij}
H. Olijnyk, A.P. Jephcoat, D.L. Novikov, N.E. Christensen, Phys. Rev. B 62, 5508 (2000).

\bibitem{goncharov}
A.F. Goncharov, and V.V. Struzhkin, Physica C 385, 117 (2003).

\bibitem{simonelli}
L. Simonelli, V. Palmisano, M. Fratini, M. Filippi, P. Parisiades, D. Lampakis, E. Liarokapis, and A. Bianconi, Physical Review B 80, 014520+ (2009), ISSN 1098-0121, URL http://dx.doi.org/10.1103/PhysRevB.80.014520.

\bibitem{15choi}
H. J. Choi, D. Roundy, H. Sun, M. L. Cohen, and S. G. Louie, Physical Review B 66, 020513+ (2002). URL http://dx.doi.org/10.1103/PhysRevB.66.020513.

\bibitem{15choi2}
H. J. Choi, D. Roundy, H. S. M. L. Cohen, and S. G. Louie, Nature (London) 418, 758 (2002).

\bibitem{15ummarino}
G. A. Ummarino, R. S. Gonnelli, S. Massidda, and A. Bianconi, Physica C: Superconductivity 407, 121 (2004), ISSN 09214534, URL http://dx.doi.org/10.1016/j.physc.2004.05.009.

\bibitem{10}
A. Bussmann-Holder and A. Bianconi, Physical Review B 67, 132509+ (2003). URL http://dx.doi.org/10.1103/PhysRevB.67.132509.


\bibitem{19}
Yu. E. Lozovik and A. A. Sokolik, Physics Letters A 374, 326 (2009), ISSN 03759601, URL http://dx.doi.org/10.1016/j.physleta.2009.10.045.

\bibitem{20}
N. Kristoffel and K. Rago (2010), 1003.5083, URL http://arxiv.org/abs/1003.5083


\end{thebibliography}
\end{document}